\documentclass[aps,prd,twocolumn,showpacs,superscriptaddress,groupedaddress,nofootinbib]{revtex4}
\usepackage{graphics}
\usepackage{epsfig}

\usepackage{graphicx}  
\usepackage{dcolumn}   
\usepackage{bm}        
\usepackage{amssymb}   
\usepackage{graphicx,color}
\usepackage{amsmath,amssymb,amscd}
\usepackage{subfigure}
\usepackage{slashed}
\usepackage{array}
\def \Lag{\mathcal L}

\newcommand*{\pd}{\partial}

\newcommand*{\ovl}{\overline}
\newcommand*{\ra}{\rightarrow}

\newcommand*{\gm}{\gamma}

\newcommand*{\half}{\frac{1}{2}}
\newcommand*{\PL}{\hat{L}}

\newcommand{\beq}{\begin{equation}}
\newcommand{\eeq}{\end{equation}}
\newcommand{\beqa}{\begin{eqnarray}}
\newcommand{\eeqa}{\end{eqnarray}}

\begin{document}
\title{ Doubly charged vector leptons and the Higgs portal}
\author{Wen Jun Li$^{1, 2 \ *}$, and
J. N. Ng$^{2}$
\vspace{5mm}\\
\normalsize\emph{  $1$ Institute of Theoretical Physics, Henan Normal
University, Xinxiang 453007, China  }\\
\normalsize\emph{$2$ Theory Department, TRIUMF, 4004 Wesbrook Mall, Vancouver, British Columbia, V6T 2A3, Canada} \\}

\email[Corresponding author:\;\;]{liwj24@163.com}
\begin{abstract}
Using a bottom up phenomenological approach we constructed a simple doubly charged vector lepton $E^{\pm\pm}$ model for the possible 750 GeV diphoton resonance $\Phi$ at the LHC assuming it to be a scalar particle. Since no stable doubly charged leptons are seen, to facilitate their decays
we complete the model by adding a charged standard model(SM) electroweak scalar $S^\pm$. $\Phi$ is a SM singlet and can be either an inert scalar or a Higgs field. In the inert case more than one vector lepton is required to
account for the photon fusion production of the resonance if the model is to remain perturbative.
For a Higgs boson case $S^\pm$ can assist the production mechanism without using more than one such lepton.
We also found that precision measurements constrain the couplings of $E^{\pm\pm}$ and $S^\pm$ to SM particles to be small. This raises the possibility that they can be fairly long lived and can give rise to
displaced vertices if produced at the LHC.
\end{abstract}

\pacs{12.60.-i,13.30.Ce}
\maketitle


A standard model(SM) singlet scalar is an important ingredient in the popular Higgs
portal \cite{PW}\cite{CNW1} scenario for dark matter models. In the simplest case
the only SM field $\Phi$ interacts with is the Higgs field $H$ via interactions
such as $\Phi^\dagger \Phi H^\dagger H$ and possibly $\Phi H^\dagger H$ depending
on whether an extra symmetry is invoked. This makes $\Phi$ very difficult to detect
both at high energy collider experiments and low energy precision measurements.
This is because effects of $\Phi$ can only arise through mixing with the Higgs
boson and this mixing is known to be small, $< 0.04$ \cite{FRU}. Hence, it is
important to explore ways to induce/enhance couplings of $\Phi$ to other SM fields.
In doing so the detection probability $\Phi$ is increased as there are more channels to explore if it is light enough to be produce at the LHC or a future circular collider. A simple possibility is to add vectorlike fermions that are charged under the SM gauge symmetries and they can couple to $\Phi$.
These fermions are vectorlike due to anomaly considerations. The simplest case is
a $SU(2)$ singlet vector lepton with $U(1)$ hypercharge $Y\neq 0$. For $Y=1$ such a lepton mixes with the righthanded SM leptons and thus leads to fine-tuning of
parameters of such a model. Moreover, for $Y\geq 2$ there is no tree level
mixing with SM leptons and this greatly simplifies the analysis of such models.
In this paper we examine the phenomenology of adding to the SM a $Y=2$ vector lepton
that carries two units of electric charge. It is denoted by $E^{\pm\pm}$. An immediate observation is that $\Phi$ can now decay into
$2 \gm$ and $ Z\gm$ via one-loop effects of the vector lepton, thus making the Higgs portal particle directly observable
at the LHC if it is sufficiently light and the parameters of the model are favorable.

In addition to the ability of enhancing the detectability of a Higgs portal scalar, $E^{\pm\pm}$ also naturally leads to lepton flavor violating
processes. We explain this assertion later. Since the SM offers no understanding
of why there are three generations of chiral fermions with masses apparently
generated by electroweak symmetry breaking, it is important to explore other avenues
in flavor physics. Doubly charged vector leptons are one such venue that
as far as we know have not being fully explored.

Recently the ATLAS and CMS collaborations reported the observation of excess events in the diphoton mass distribution around 750 GeV \cite{Atlas},\cite{CMS} in the $\sqrt s=13 $ TeV Run II data recorded with $3.2 \,\mathrm{fb}^{-1}$ of pp collisions. This unexpected development has understandably generated a great deal of interest among theorists. It is common to interpret this in terms of a new spin-0 resonance, although a spin-2 particle is not ruled out. It is also noted that the same is not seen in the dijet mass spectrum. On the other hand, the event rate and lack of signal at lower energies appear to favor a scalar or pseudoscalar resonance being produced and decays predominantly into two photons. A plausible explanation is that this resonance couples predominantly to two photons and the couplings to gluons and other colored objects are suppressed or do not exist at all. This was first discussed in
\cite{FGR},\cite{CHT}. In this case, the resonance is produced via two photon fusion and both exclusive and inclusive processes can take place with the latter being more important. It can also be shown that other fusion mechanisms such as photon Z and two Z bosons are less important. More recent LHC Run II data from both CMS \cite{CMS2016} and
Atlas \cite{Atlas2016} do not support the initial data. With  $12.9 \, \mathrm{fb}^{-1}$ pp collisions collected no new signal above background was recorded. This reduces the local significance of the
initial excess from $\sim 3.6$ to $\sim 2.3 \sigma$.

In this paper we take the data at face value and interpret the combined results of 2015 and 2016 data as an
upper limit on the photon fusion production of a spin-0 resonance of mass 750 GeV. This is to be taken as an example of the limits on the parameters of the model we discuss later. As reported by Atlas, two of the 15 original excess events
are consistent with background; and  no new excess is found in the 2016 run. Thus far we estimate this cross section to be $\lesssim 0.84$ fb \cite{FGR}. Focusing on the case of a scalar $\phi$, the effective Lagrangian for the above process is given by
\beq
\label{eq:lgg}
\Lag = \frac{1}{f_\gm}\phi (F_{\mu\nu})^2,
\eeq
and $f_\gm \gtrsim 8.6 $ TeV \cite{FGR}. Since Eq.(\ref{eq:lgg}) can only come from a one-loop effect this sets a limit on the coupling of $\phi$ to the particles in the loop .

In this paper we take $\phi$ to be a SM singlet scalar field and identify it as the real part of the complex Higgs portal field discussed previously. We study the possibility
that it is a bridge to new vector leptons $E$ of hypercharge $Y=2$ or higher, which are also $SU(2)$ singlets. We do not extend the gauge symmetry of the SM and the vector nature of these leptons
 does not lead to anomalies. The number of such particles is not known. The physics we wish to explore is well illustrated by considering just
one such particle. Extending to more is straightforward. Similarly higher charged particles can also
be considered. For simplicity we take $Y=2$ and comment on other possibilities when appropriate.
Since there are no stable doubly charged leptons, it is mandatory that $E$ has decay channels.
Charge and angular momentum considerations dictate the decay to be $E^{--}\ra \ell+ \ell^\prime +\nu^{c}$ where $\ell,\ell^\prime =e,\mu,\tau$ and $\nu$ is an active SM neutrino.
The flavor of $\nu$ is different from $\ell$ and/or $\ell^\prime$.\footnote{$E^{--}\ra
\ell+W^-$ is allowed by charge and angular momentum considerations but forbidden by SM gauge symmetries.} An explicit example is $E^{--}\ra e^- \mu^- \nu_\tau^{c}$. If we assign unit lepton number to $E^{--}$ the decay conserves global lepton number; however,lepton flavor is violated.  The scale of lepton flavor violation is given by the
mass of the mediating particle that gives rise to the above decay. No SM fields can lead to the above decay. The simplest solution is to introduce a $Y=1$ $SU(2)$ singlet scalar $S$. If $S$ is lighter than $E$ then the decay is sequential: $E\ra S +\ell $ followed by $S\ra \ell^\prime + \nu^c$. On the other hand,
if $E$ is lighter than $S$, the decay is a three-body mode similar to that of ordinary muon decays.

The quantum numbers of the new particles together with the relevant SM fields are given in Table I below
\begin{table}
\label{tb}
\caption{\small{ Quantum numbers of the SM Higgs $H$, leptons $L,\ell$ and $E,S,\phi$}}
\begin{center}
\renewcommand{\arraystretch}{1.30}
\begin{tabular}{|c|c|c|}
\hline
Field&$SU(2)$&$U(1)_Y$\\ \hline
$H$ & {\bf{2}}& $\phantom{-}\frac{1}{2}$ \\ \hline
$L$ & {\bf {2}} & $-\frac{1}{2}$ \\ \hline
$\ell_R$ & {\bf{1}} &$-1$\\ \hline
$E$ & {\bf{1}}& $-2 $\\ \hline
$S$ & {\bf{1}}& \phantom{-}1 \\ \hline
$\Phi$ &{\bf{1}}&\phantom{-}0\\ \hline
\end{tabular}
\end{center}
\end{table}
where standard notations are used.

In addition to the SM Lagrangian that involving the new fields is given by
\beqa
\begin{split}
\label{eq:Lag}
&{\Lag}^{\prime} \\
=&\ovl{E} i\gamma^\mu (\pd_\mu -2ig_1 B_\mu) E +[(\pd^\mu +i g_1 B^\mu)S]^\dagger (\pd_\mu +i g_1 B_\mu)S \\
&-\left[f_{e\mu}(\ovl{\nu_e^{c}} \mu_L -\ovl{\nu_{\mu}^c}e_L)+f_{e\tau}(\ovl{\nu_e^{c}}\tau_L  -\ovl{\nu_{\tau}^c}e_L) \right.\\&\left.+f_{\mu\tau}(\ovl{\nu_{\mu}^c}\tau_L -\ovl{\nu_{\tau}^c}\mu_L)\right]S  -y_E \ovl{E} E\Phi -M_E \ovl{E}E
 \\
&-\sum_a^{e,\mu,\tau} y_a\ovl{E}\ell_{Ra} S^\dagger -V(H,S,\Phi) + h.c.,
\end{split}
\eeqa
 the scalar potential $V(H,S,\Phi)$ is
\beqa
\label{eq:scalarV}
\begin{split}
V&=-\mu^2 H^\dagger H +\lambda (H^\dagger H)^2 +M_S^2 S^\dagger S + \lambda_S (S^\dagger S)^2 \\
&+\lambda_{SH} S^\dagger S H^\dagger H +\lambda_\phi (\Phi^\dagger \Phi)^2 +M_{\phi}^2\Phi^\dagger \Phi +\lambda_{\phi h}\Phi^\dagger \Phi H^\dagger H \\
&+
\lambda_{\phi S}\Phi^\dagger\Phi S^\dagger S +\alpha \Phi +\beta \Phi^\dagger \Phi \Phi +\kappa_H \Phi H^\dagger H \\
&+\kappa_S \Phi S^\dagger S.
\end{split}
\eeqa

 In general, $\Phi$ can be complex. For simplicity we take $\Phi$ to be real.
  The imaginary part plays no role in what we study here since we
 take $\phi$ in Eq.(\ref{eq:lgg}) to be a scalar. The usual Higgs portal potential can be obtained from Eq.(\ref{eq:scalarV}) by deleting the $S$ field. How $\Phi$
 connects to the dark matter is model dependent and is not pursued here.
Note that lepton number is conserved in this model and neutrinos remain massless. In order to give masses to active neutrinos, one can implement type I seesaw by adding heavy singlet neutrinos or radiatively generating them by adding a second scalar
doublet as in the Zee model \cite{Zee}. Though interesting we do not pursue this further here.

Taking the U-gauge for the Higgs doublet, we parametrize $H$ and $\Phi$ by
\beq
\label{eq:hphi}
H=\begin{pmatrix}0\\ \frac{v+h}{\sqrt{2}}\end{pmatrix}, \quad \Phi=\frac{w+\phi}{\sqrt{2}},
\eeq
where $v,w$ are the respective vacuum expection values(VeV) of $H$ and $\Phi$ fields. The stationary conditions for $H,\Phi$ are
\beqa
v \left(-\mu^2 +\lambda v^2 +\frac{ \lambda_{\phi h} w^2}{2} +\frac{\kappa_H w}{\sqrt {2}}\right)&=&0 ,\nonumber\\
w \left(M_{\phi}^2 +\lambda_\phi w^2 +\frac{ \lambda_{\phi h} v^2}{2} +\frac{\alpha}{\sqrt{2}w}+\frac{3\beta w}{2\sqrt{2}}+\frac{\kappa_H v^2}{2\sqrt{2}w}\right)&=&0.
\nonumber \\
\eeqa
If $w=0$ then $\Phi$ is not in the  Higgs phase. However, this requires $\alpha+\kappa_H v^2/2=0$.

 A second possibility is $w\neq 0$ and $\Phi$ is also a Higgs field. The stationary condition can easily be satisfied for $M_\phi ^2 <0$ although this is not the only possibility. The trilinear terms $\phi h h$
and $\phi S^+ S^-$ are present whether $\Phi$ is in the Higgs phase or not. Furthermore, $h$ and $\phi$ in general mix.

If $w=0$ then their mass square
matrix of $(h,\phi)$ is expressed in
\beq
\label{eq:massmatrix}
\half \begin{pmatrix}h&\phi\end{pmatrix}\,\begin{pmatrix}2v^2\lambda  & \frac{v\kappa_H}{\sqrt{2}} \\ \frac{v\kappa_H}{\sqrt{2}} & \bar{M_\phi}^2 \end{pmatrix} \begin{pmatrix}h\\ \phi \end{pmatrix},
\eeq
where $\bar{M_\phi}^2 =M_\phi^2 +\lambda_{\phi h} v^2/2$. $(h,\phi)$ is related to the mass eigenstates $(h^\prime,\phi^\prime)$ by the usual $2\times 2$
rotation matrix defined by the mixing angle $\theta$ that is given by
\beq
\label{eq:mixing}
\tan {2\theta} =\frac{\sqrt{2}v \kappa_H}{\bar{M_\phi}^2 -v^2 \lambda}.
\eeq
If we identify $\phi^\prime $ as the 750 GeV  resonance and $h^\prime$ as the SM-like Higgs
with mass 125 GeV it is natural to assume $M_\phi>v$. In the limit $\kappa_ H\ra 0$
the two fields decouple from each other. The observed Higgs boson is SM like and the
mixing with another scalar is limited by $\sin^2 \theta \lesssim 0.04$ \cite{FRU} from an analysis
of Higgs coupling strength data from LHC Run-I. We also note that another analysis \cite{NRM} gives
a larger value of $\sin^2 \theta \lesssim 0.33$. Using the more stringent constraint we estimate that $\frac{\kappa_H}{v}\lesssim 5$. It is interesting that current data allow $\kappa_H$ to be $O(\mathrm{TeV})$. It can be much smaller if the data on SM Higgs couplings become more stringent.\footnote{Since $\phi\ra hh$ is allowed and we require that it does not dominate over the diphoton mode. We then obtain the constraint $\frac{\kappa_H}{M_\phi} \lesssim O(\frac{\alpha_{EM} y_E}{4\pi})$. This  becomes clear later.}  On the other hand $\kappa_S$ remains unconstrained.

For notational simplicity we drop the prime in the mass eigenstates.

For $w\neq 0$ the neutral scalar mass square matrix is more complicated. It can be obtained from
Eq.(\ref{eq:massmatrix}) by the following substitutions: $\bar{M_\phi}^2\ra \lambda_\phi w^2 -\frac{\alpha }{\sqrt{8}w}+\frac{3\beta}{\sqrt{8}}-\frac{\kappa_H v^2}{\sqrt{32}w}$ and
$\frac{\kappa_H v}{\sqrt{2}}\ra \lambda_{\phi h}vw+\frac{\kappa_H v}{\sqrt{2}}$. The mixing is given by Eq.(\ref{eq:mixing}) with the above substitutions.

The mass parameters $M_E,M_S,M_\phi$ in Eq.(\ref{eq:scalarV}) are all free parameters. The physical masses depends on
whether $\Phi$ is in the Higgs phase or not. The mass of $E$ is just $M_E$ since there is no mixing with the SM leptons. For $w=0$,
the mass S is given by $M_S^2+\lambda_{SH}v^2/2$. Similarly the mass of $\phi$ is approximately
given by $M_\phi ^2 +\lambda_{\phi h}v^2/2$. The relative size of these masses cannot
be determined. The case of Higgs boson $\Phi$ is similar with more complicated formulas.
For definiteness we take $E$ to be heavier than $S$ and the physical mass of $S$ is greater
than 80 GeV from LEPII searches \cite{LEPII}.

With the Lagrangian in place the effective Lagrangian Eq.(\ref{eq:lgg}) is obtained from
calculating Fig.1.
\begin{figure}[ht]
\centering
\subfigure{
\includegraphics[width=3cm]{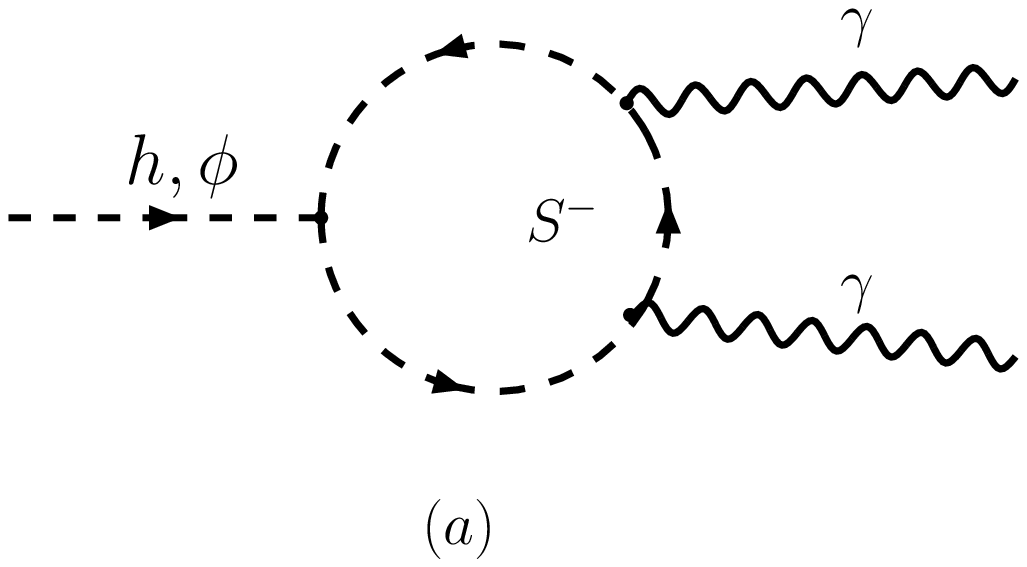}}
\;
\subfigure{
\includegraphics[width=3cm]{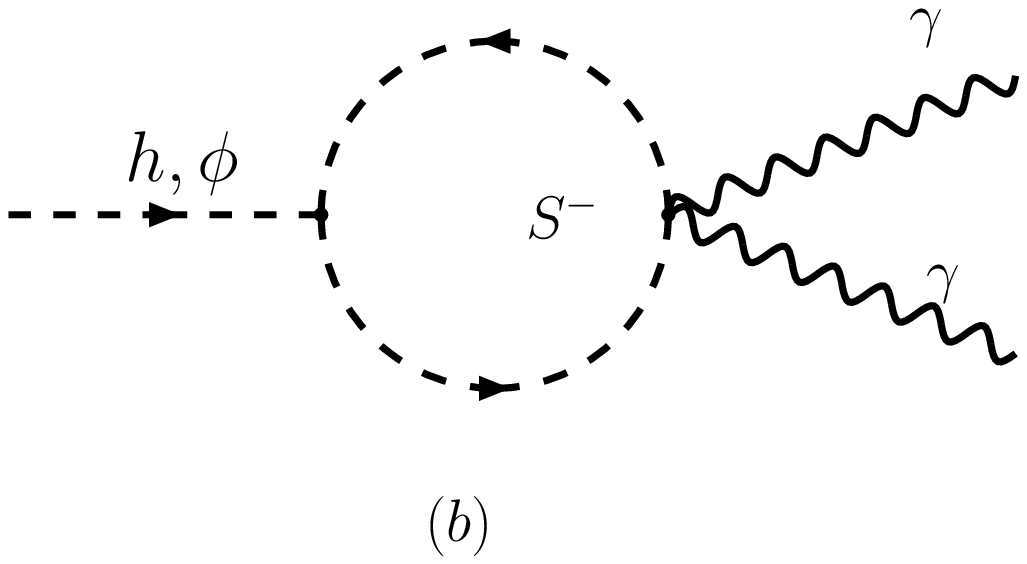}}
\;
\subfigure{
\includegraphics[width=3cm]{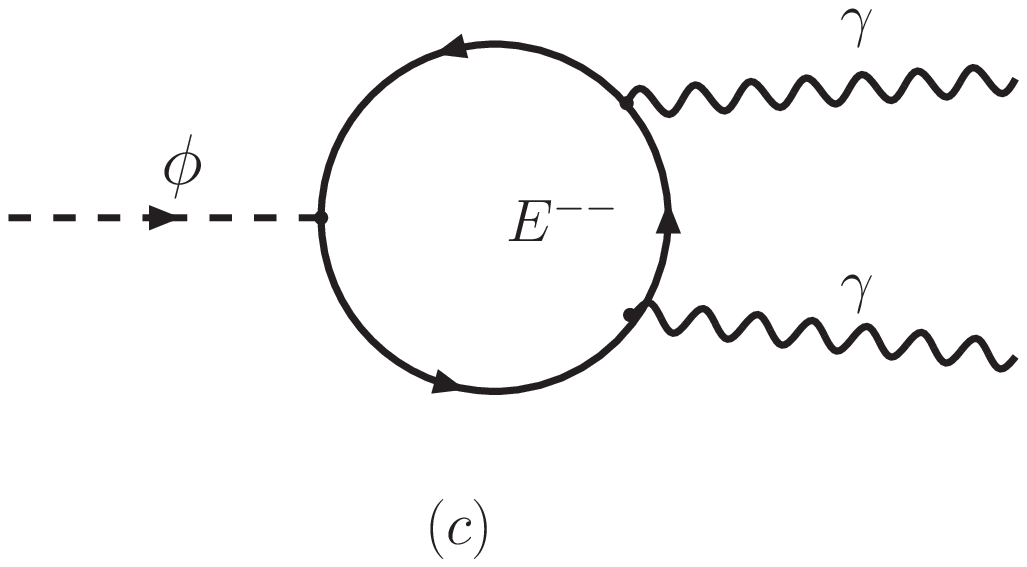}}
\caption{ (a) and (b) $\phi$ to two photons via the charged $S$ scalar loop for $\Phi$ in the Higgs phase. They also contribute to SM Higgs decays. (c) The $E$ loop contribution.}
\label{fig:lgg}
\end{figure}

All three diagrams contribute to $\phi$ decays whether they are Higgs boson or not. For the same SM Higgs decays and neglecting the small $\phi-h$ mixing the $S$ loop contributes but the $E$ loop does not. The calculation of the above diagrams gives \cite{CNW}
\beq
\label{eq:fgamma}
\begin{split}
&f_{\gm}^{-1}=\\&\frac{\alpha}{ 4\pi M_\phi}\left(Q^2 N y_E\sqrt{\tau_E}F_{\half}(\tau_E)+
\frac{2(\lambda_{\phi S}w+\kappa_S)}{M_S}\sqrt{\tau_s} F_0(\tau_s)\right).\end{split}
\eeq
We have used the convention of \cite{tome} and define $\tau_i= M_\phi^2/(4M_i^{2})$, and the one-loop
functions are
\beqa
F_0(\tau)&=&-[\tau-f(\tau)]\tau^{-2}, \\
F_\half(\tau)&=& 2[\tau+(\tau-1)f(\tau)]\tau^{-2},
\eeqa
with
\beq
f(\tau)=
\begin{cases}
\arcsin^2{\sqrt{\tau}} & \tau\leq 1 \\
-\frac{1}{4}\left[\log{\frac{1+\sqrt{1-\tau^{-1}}}{1-\sqrt{1-\tau^{-1}}}} -i\pi\right]^2 & \tau > 1.
\end{cases}
\eeq
From the event rates given one can deduce that
$f_\gm \sim 8-9$ TeV \cite{FGR}. Eq.(\ref{eq:fgamma}) gives strong constraints on the model
parameters since the $F$ functions are known. We plot them in Fig.2.
\begin{figure}[t]
\begin{minipage}{0.5\linewidth}
\hskip-1.5cm
      \includegraphics[width=2.7in]{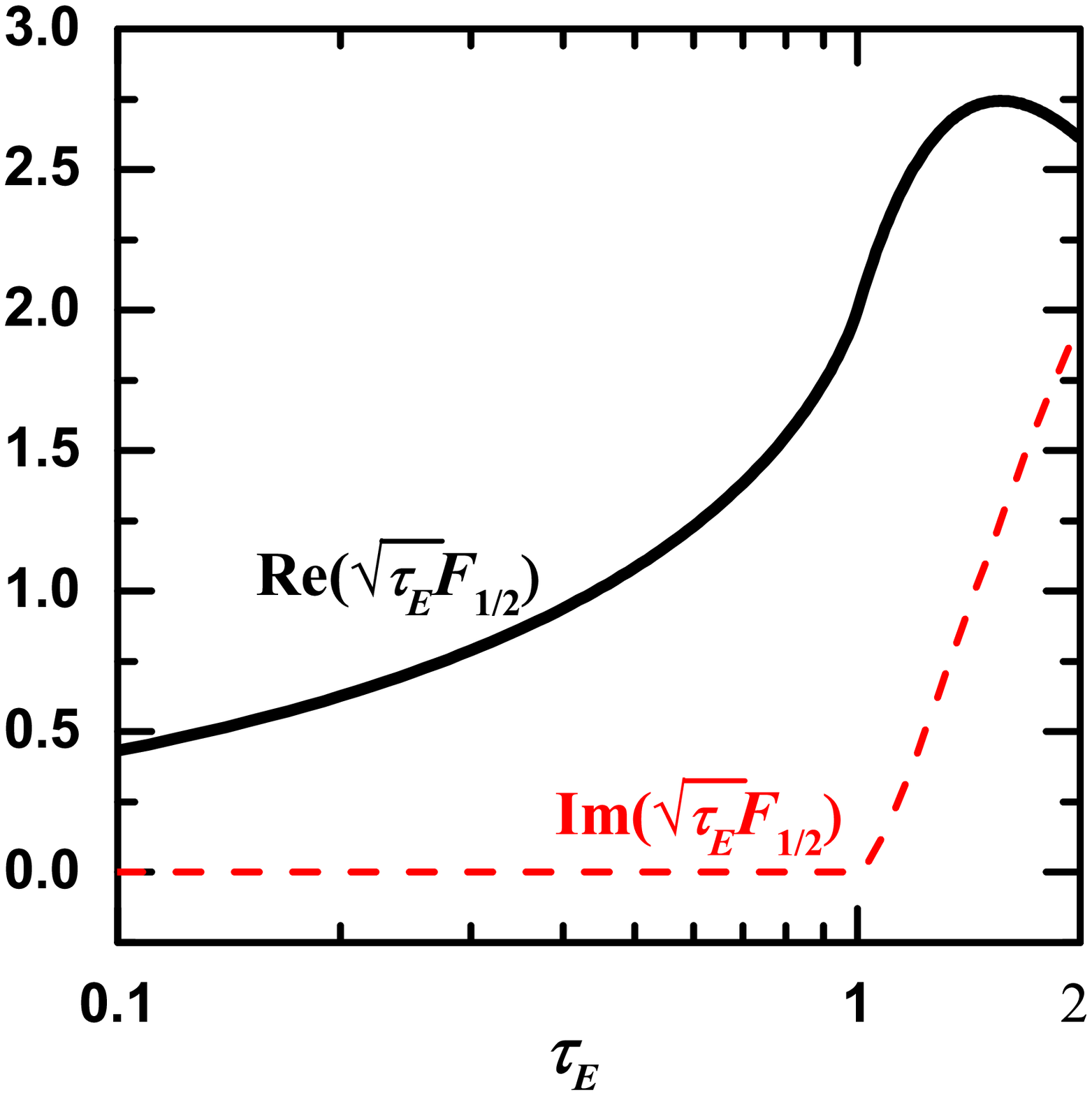}
      \centering{(a)}
  \end{minipage}%
\begin{minipage}{0.5\linewidth}
      \hskip-1cm
      \includegraphics[width=2.7in]{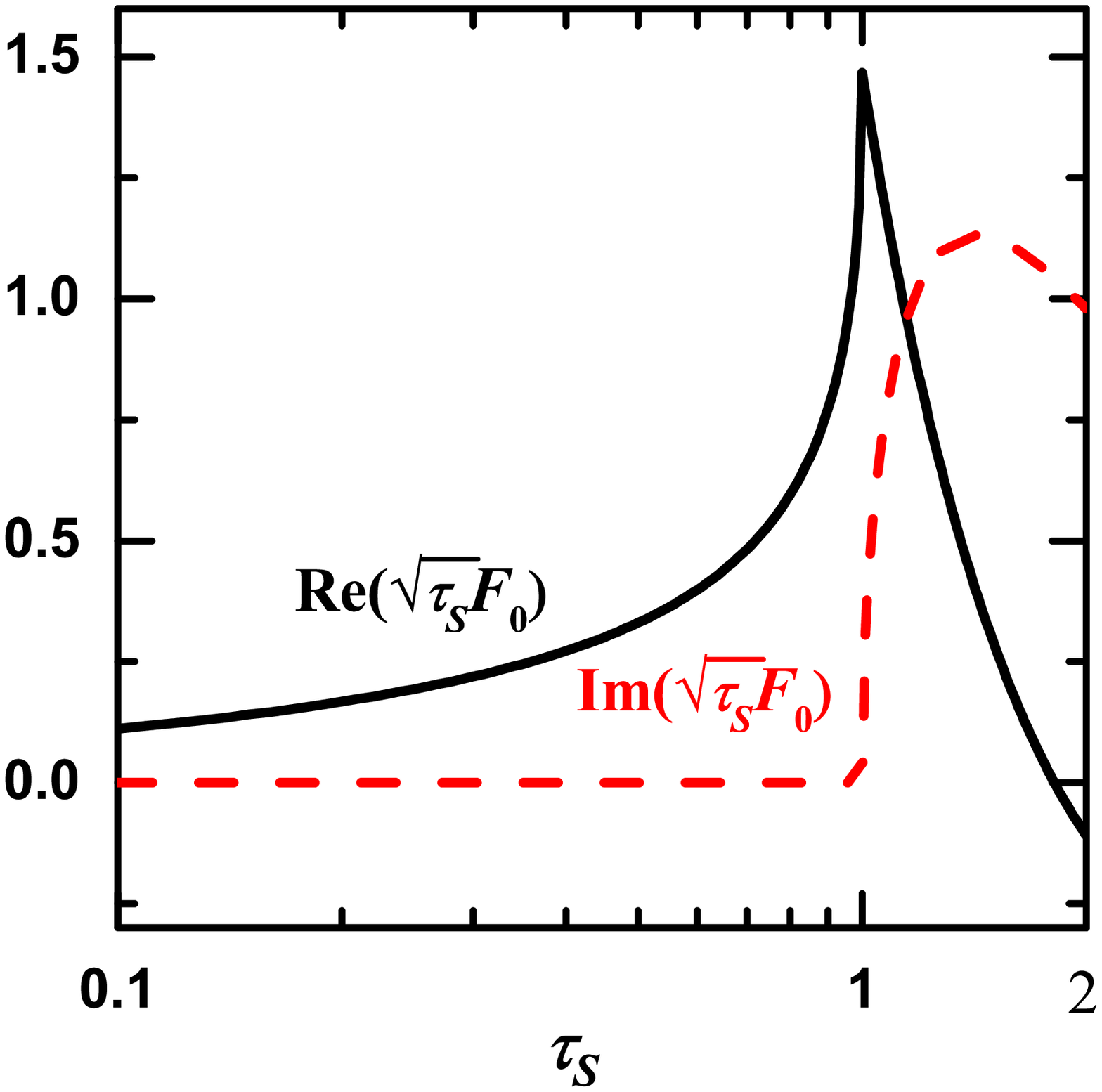}
 \centering{(b)}
  \end{minipage}%
  \caption{Form factors for (a) spin-$\half$ and (b) spin-0 particle contributions to $\gm\gm$
couplings to a scalar as a function of $\tau_i=M_\phi^2/4M_i^2$ with $M_i$ mass of the loop particle.}
\label{fig:FF}
\end{figure}

First we consider that $\Phi$ is not a Higgs scalar, i.e., $w=0$. Since there is no constraint on the value of $\kappa_S$ we consider two limits.
\begin{enumerate}
\item
We first take it to be small, i.e., $\kappa_S<<v$.  Then the E loop has to account for the observed events. For the simplest case of one E we have $Q=2,N=1$. The function $\sqrt{\tau}F_\half(\tau)$ has the value of 2.0 at $\tau= 1$ and falls rapidly for $\tau<1$. As a benchmark point we take $f_\gm = 9 $ TeV and obtain the constraint
\beq
y_E\ N\simeq 16.8.
\eeq
Clearly for $N=1$ the Yukawa coupling is so large as to invalidate perturbative calculations. In order for  $y_E$ to reside in the perturbative regime, i.e., $<4 \pi$, it is required that $N>2$.
In the region of $\tau_E <1$, $\phi$ does not decay into $E\bar{E}$ pairs and hence does not
lower the two photon branching ratio.

For the region $1<\tau_E <2$, we have $Re(\sqrt{\tau}\,(F_\half))$ falling slowly and the imaginary part rising very fast by comparison. Taking the peak value of $\sqrt{\tau}F_\half =2.5$ still requires $y_E\sim O(1)$ unless $N>10$. Barring the very high multiplicity case, $y_E \sim O(1)$ implies that the dominant decay of $\phi$ is into $E\bar{E}$ pairs  instead of the two photons mode.

 An alternative to adding more doubly charged leptons is to add a higher charged vector lepton. For example, we can add an $E^{\pm\pm\pm}$. This has the same effect as adding two $E^{\pm\pm}$ assuming that their Yukawa couplings are not too different.
 We note that the triply charged $E$ is also unstable and can decay via $E^{---}\ra E^{--}+S^-$.
\item Next we take $\kappa_S\gtrsim v$. We have seen that $\kappa_H \sim $ TeV is allowed by the mixing data; perhaps this not an unreasonable domain for $\kappa_S$ to be in. The scalar
    loop contribution is given by $F_0$; see Eq.(\ref{eq:fgamma}). It is smaller than $F_\half$ by a factor of 2 for the same value of the arguments in the region $\tau<1$.
 This can be compensated by adjusting $\kappa_S/M_S$ and we set $w=0$ for now. The constraint is given by
 \beq
 Ny_E \sqrt{\tau_E}F_\half(\tau_E) +\frac{\kappa_S}{2M_S}\sqrt{\tau_s}F_0(\tau_s)\lesssim 34.
 \eeq

To have a significant effect, we
 require $\kappa_S \gtrsim 10$ TeV. For $\kappa_S$ in the $\gtrapprox 20$ TeV
range the scalar loop dominates and all couplings can remain perturbative even for
 $N=1$.
\end{enumerate}

The same considerations can be applied to the case of a Higgs boson $\Phi$. The main difference is that
 we require $\lambda_{\phi S}w +\kappa_S$ to be in the 10 TeV range; and the scalar loop dominates and all couplings can remain perturbative. To give an example, let $
 w(\kappa_S)=10(15)$ TeV, $M_S=400$ GeV, $M_E=500$ GeV; we obtain
 $y_E=2.6$ for $\lambda_{\phi S}=1$ and $N=1$.  Moreover,
 for large values of $w$ a mild fine-tuning
  of $\lambda_\phi$ is required to get a 750 GeV scalar. There is a further fine-tuning problem with this solution. The physical mass of $S^\pm$ is given by
 \beq
 \label{eq:msphy}
 M^2_{S}(phy)= M_S^2 +\frac{\lambda_{SH}v^2}{2}+\frac{\lambda_{\phi S}w^2}{2}+\frac{\kappa_S w}{\sqrt{2}}.
 \eeq
 The sum of the first two terms has to be negative and large in order to
 provide cancelation to the large $\kappa_S$ and $w$ contributions, i.e., $O(10)$ TeV, so that $M_S(phy)\sim 0.4$ TeV. It is easy to see from Eq.(\ref{eq:scalarV}) that the extremum condition on $S$ can be written as $M_S^2(phy)+2\lambda_S S^\dagger S$ at the electroweak and singlet minimum and is positive definite. Hence, there is no charge breaking vacuum here.

In passing we note that the region of $\tau_s>1$ is ruled out since $\phi\ra S^+S^-$ is the dominant decay.

 In either phase of the $\Phi$ field, both $E$ and $S$ have to be heavier than $M_\phi/2$ in order
 for the model to accommodate the current data for the 750 resonance. However, which one is heavier cannot be determined yet.

 The model we constructed with new fields carrying only $U(1)_Y$ quantum numbers leads to the
 prediction that the ratios of widths into SM gauge bosons are \cite{LL}
 \beq
 \begin{split}
 \Gamma_{\gm\gm}:\Gamma_{\gm Z}:\Gamma_{Z Z}&= 1: \frac{2s_w^2}{c_w^2} : \frac{s_w^4}{c_w^4}\\
 &\approx 1:0.54:0.07 \\
 \Gamma_{W W}&=0,
 \end{split}
 \eeq
where $s_w(c_w)$ is the sine(cosine) of the weak mixing angle and in the limit that $\kappa_H$ is small.

An important consideration of introducing new heavy charged states is to
examine the constraints of low energy precision measurements put on their masses and couplings.
If $E^{\pm\pm}$ and $S^\pm$ were to play roles in the diphoton resonance as we discussed above,
then both masses have to be $> 375$ GeV. This is higher than the constraints imposed from the model independent bound  from LEP II since no doubly charged leptons or charged scalars were seen \cite{PDG1}. Closer examination of Fig.2 reveals that if $M_S$
stays close to 375 GeV it offers the most impact while $M_E$ can be larger due to the
slower falloff of $Re (\sqrt{\tau}F_\half)$.
As stated before  we take the vector lepton to be the heavier one and use $M_S=400\, \mathrm{GeV}$ as a benchmark. For the opposite case of $M_S> M_E$ there is no
qualitative difference. Quantitatively since $F_0$ falls off very fast as
$M_s$ increases, a larger $\kappa_S$ is required.

Next we examine the constraints from low energy physics. The exchange of $S$ in muon decays modifies the Fermi coupling $G_F$ as measured by muon lifetime. With new physics in the leptonic sector we assume instead unitarity of quark mixing and extract the $G_F$ from nuclear, kaon, and B-meson decays \cite{GF}. This gives $G_F= 1.166309(350)
\times 10^{-5}\mathrm{GeV}^{-2}$. The effective Lagrangian due to $S$ exchange yields
\beq
\label{eq:LS}
\Lag=\frac{if_{e\mu}^2}{2M_S^{2}}\left(\ovl{\nu_\mu} \gamma^\alpha \PL \nu_e \right)\left(\bar{e}\gamma_\alpha \PL \mu\right),
\eeq
whereas the SM has $-\frac{ig^2}{2M_W^{2}}$ in front of the four-Fermi operator. Here $\PL =(1-\gamma_5)/2$. Thus, we obtain
\beq
\label{eq:femu}
f_{e\mu}\leq 1.502 \times 10^{-1} \left(\frac{M_S}{400\mathrm{GeV}}\right).
\eeq\

Similarly using the leptonic $\tau$ decay ratio into $\mu,e$, we get
\beq
\begin{split}
\frac{\Gamma(\tau\ra \mu+\nu's)}{\Gamma(\tau\ra e+\nu's)}&=\frac{\left(1-\frac{f_{\mu\tau}^2M_W^2}{g^2M_S^2}\right)^2+\cdots}{\left(1-\frac{f_{e\tau}^2M_W^2}{g^2M_S^2}
\right)^2+\cdots}\\
&\simeq 1+2(f_{e\tau}^2-f_{\mu\tau}^2)\left(\frac{M_W^2}{g^2M_S^2}\right),
\end{split}
\eeq
where $\cdots$ denotes terms such as $f_{\mu e}^2f_{\tau e}^2$, which come from diagrams that interfere
incoherently with the SM ones. They are of order $f^4$, which we neglect. Thus,
\beq
\label{eq:taud}
f_{e\tau}^2-f_{\mu\tau}^2\leq \pm 2.25 \times 10^{-2} \left(\frac{M_S}{400\mathrm{GeV}}\right)^2,
\eeq
where the experimental value of $\frac{\Gamma(\tau\ra\mu^- \bar{\nu_\mu}\nu_\tau)}{\Gamma(\tau\ra e^- \bar{\nu_e}\nu_\tau)}=0.979\pm 0.004$ has been used \cite{PDGtau}.

Next we consider the rare decay of $\mu\ra e\gm$. The Feynman diagrams to calculate are
given by Fig.3.
\begin{figure}[ht]
\centering
\includegraphics[width=9. cm]{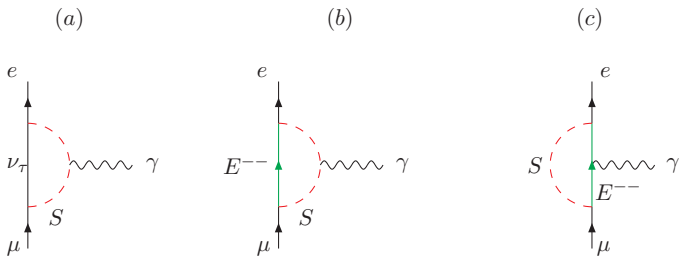}
\caption{Diagrams leading to $\mu\ra e\gm$.
Wave function renormalization graphs are not shown}
\label{fig:mueg}
\end{figure}
We have set $m_e=0$ and note that diagram (a) has a different chiral structure than (b) and (c) and they add incoherently.
Using the experimental bound of $BR(\mu\ra e \gm)< 5.7\times 10^{-13}$ \cite{mueg} we get the strong constraint
\beq
\begin{split}
\label{eq:mueg}
&f_{e\tau}^2 f_{\mu\tau}^2 +\left(\frac{y_e y_\mu x^2}{(1-x)^4}\right)^2\left[(-4+9x-5x^3)+6x(2x-1)\ln x\right]^2 \\
&\le 2.235 \times 10^{-12}\left(\frac{M_s}{400\mathrm{GeV}}\right)^4,
\end{split}
\eeq
where $x=\frac{M_S^2}{M_E^2}$. Assuming that the two terms on the left-hand side are of the same order then $f_{e\tau}f_{\mu\tau}\lesssim 10^{-3}$. For $x<1$ the terms multiplying $y_e y_\mu$ can give a large coefficient, e.g., for $x=0.64$ this factor is $\sim 18.18$. This implies that $y_e y_\mu$ is about the same order of magnitude as that of the $f$'s. Certainly it can be smaller in which case the $f_{e\mu}f_{\mu\tau}$ saturates the bound.

Similar diagrams with the final electron state replaced by a muon contribute
to muon anomalous moment $a_\mu$. This contribution is
\beq
\label{eq:g-2}
\begin{split}
\Delta a_\mu &=\frac{m_\mu^2}{96\pi^2 M_S^2}\biggl((f_{\mu\tau}^2+f_{e\mu}^2)\\&-\frac{y_\mu^2 x}{(1-x)^4}\left[4-9x+5x^3-6x(2x-1)\ln x\right]\biggl).
\end{split}\eeq
Putting in the numbers we get
\beq
\begin{split}
f_{\mu\tau}^2+f_{e\mu}^2
&-\frac{y_\mu^2 x}{(1-x)^4}\left[4-9x+5x^3-6x(2x-1)\ln x\right]\\
&\le 39.08\left(\frac{M_S}{\mathrm{400}}\right)^2,
\end{split}
\eeq
where we have used $a_\mu^{exp}-a_\mu^{SM}=2.88\times 10^{-9}$ \cite{mug2}.
The above considerations constrain most of the parameters of Eq.(\ref{eq:Lag}).

On the other hand, there are fewer limits on the
parameters of the $V(H,S,\Phi)$. Since the potential preserves custodial symmetry of the SM, most
electroweak precision measurements take the SM values. Moreover, from Fig.1
 we see that the scalar loop can modify the $h\ra \gm\gm$ signal compared to the SM. In
contrast the vector lepton makes no contribution at this level. Experimentally the signal strength of $h\ra\gm\gm$ is very close to the SM value we can use to constrain $\lambda_{SH}$. Defining $R \equiv\Gamma^{new}/\Gamma^{SM}$ we obtain
\beq
R=\Big|1+\frac{\lambda_{SH}v^2}{2M_S^2}\frac{F_0(\tau^\prime)}{F_1(\tau_w)+\frac{4}{3}F_\half(\tau_t)}\Big|^2
\eeq
where $\tau^\prime=M_h^2/4M_S^2$ and $ F_1(\tau)=-[2\tau^2 +3\tau +3(2\tau -1)f(\tau)]\tau^{-2}$.
The current bound on R is $1.17\pm 0.27$ \cite{Atlasmu}. This yields the constraint $|\lambda_{SH}|< 8.1$.

The above can be used to estimate the lifetimes of $E$ and $S$, since we are interested in the case of $E$ being heavier than $S$, which is the favored region from the perturbative viewpoint. The main decays are $S\ra \ell + \nu$ and $E\ra S+ \ell$. The results are
\beq
\label{eq:Swidth}
\Gamma_S=\frac{M_S}{8\pi}\sum_{\ell^\prime\neq\ell}\sum_{\ell}|f_{\ell\ell^\prime}|^2 ,
\eeq
and
\beq
\label{eq:Ewidth}
\Gamma_E=\frac{M_E}{32\pi}\left(1-\frac{M_S^2}{M_E^2}\right)^2\sum_\ell |y_l|^2,
\eeq
where the light lepton masses are all neglected. From Eqs.(\ref{eq:femu}), (\ref{eq:taud}), and (\ref{eq:mueg}) we expect $f_{\ell \ell^\prime}\lesssim  10^{-3}$ [see discussions following Eq.(\ref{eq:mueg})].
For a 400 GeV $S$ its lifetime is $ 1.4\times 10^{-20}$ sec with the assumption that $f_{e\mu}\sim f_{e\tau}\sim f_{\mu\tau}\sim10^{-3}$. For smaller $f$'s, e.g., $\sim 10^{-5}$, the lifetime is long enough to give displaced vertices, which
can aid in its detection when they are produced at the LHC. The lifetime for a 500 GeV vector lepton
decaying to a 400 GeV $S$ is estimated to be $\sim 3.4 \times 10^{-19}$ sec.
These lifetime estimates are important for finding signatures for the production of these new particles. At the LHC production of $S$ proceeds via quark antiquark annihilation at the parton level
\beq
\label{eq:sprod}
q+\bar{q}\ra S^+ S^-\ra \ell^+ \nu \ell^\prime \nu^c
\eeq
The signature is two lepton pairs that need not  have the same flavor and missing transverse
energy, $\slashed{E}_T$, with no associated jets. However, the background from W boson pair
production is severe. On the other hand, at an $e^+ e^-$ collider the signals for
$S^+S^-$ are much easier to unravel.

Similarly the production of $E$
can be searched for by the sequence of reactions
\beq
\label{eq:Eprod}
q+\bar{q}\ra E+\bar{E} \ra S^- +\ell_a^- + S^+ + \ell_b^{+}\ra \ell_a^- + \ell_c^- +\ell_b^+ + \ell_d^+ + \slashed{E}_T
\eeq
where $a,b,c,d$ denote the flavors of the charged leptons. Here the signal is four leptons plus $\slashed{E}_T$ with no associated jets. Furthermore, the charged leptons do not form invariant mass peaks. Interestingly if the couplings $f$'s and $y$'s are very small, i.e., $<O(10^{-6})$, we have displaced vertices as discussed before. As an example we take $y_e\simeq y_\mu\simeq y_\tau \sim
5\times 10^{-6}$, which are values near the experimental limits; then the production of $E\bar{E}$ pairs leaves two $\sim 2 \mathrm{mm}$ tracks from the collision point. Each subsequently leads to two more tracks depending on
the decay modes. For smaller values of the $y$'s longer tracks are expected. For a discussion of the displaced vertices search, see for example \cite{DV}. We have here the unusual case in which the LHC can cover very small couplings
that precision measurements cannot reach in the foreseeable future.

Similar to the case of $S$ high energy $e^+ e^-$ colliders offer cleaner signatures, aside from the ratio of total cross sections to muon pair production R, which gives 1 for $S^+S^-$ and 4 for $E^{++}E^{--}$, respectively. If their production is way above threshold they can give spectacular signatures with a pair of same sign leptons going in
one direction and a pair of antileptons going in the opposite direction.
 Looking for similar signatures at the LHC is more complicated. At the parton level, see Eq.(\ref{eq:Eprod}), in the parton center of mass frame, the charged lepton pair and antilepton pair emerge in opposite directions. However, since the quarks and antiquarks have different parton distribution functions, in the
laboratory frame they are boosted differently. Nevertheless, we can expect that leptons and antileptons are still separated in rapidity. We defer a detail study to a later investigation. A recent study of the Drell-Yan production of new particles related to the diphoton resonance is given in \cite{DYdi}.

In conclusion, we employed a totally phenomenological approach to construct and study a simple model of doubly charged vector leptons $E^{\pm\pm}$ that may enhance the photon fusion production of the singlet Higgs boson portal at the LHC. We used a 750 GeV diphoton scalar resonance as an example to evaluate the parameters of the model
that can lead to its detectability. The coupling $y_E$ has to be $O(1)$. Furthermore, these vector leptons are phenomenologically interesting in their own rights
and have been discussed in the context of excited leptons; see, e.g., \cite{exlep}. Moreover, they have to be unstable and thus must decay. A simple electroweak singlet charged scalar $S$ is utilized to complete the model. It is found that if $\Phi$ is not in the Higgs phase one would require two or more vector leptons in order for the model to be amendable to perturbative treatments and account for the data if the $\kappa_S$ term is small. On the other hand if $\kappa_S $ is large  then $S$ can assist
the lepton loop in giving a large enough effective coupling to accommodate the observed signal strength with only a single vector lepton required. The Yukawa couplings of $\Phi$ to these particles are of order 1 for the reference kinematic point we use. Given the preliminary nature of the data we did not pursue a detailed parameters scan. This mechanism can be carried over to the case of a Higgs boson $\Phi$. Moreover, there is a price to be paid here in that fine-tuning
of parameters in the scalar potential is needed in order to keep $S^\pm$ relatively light
and the singlet VeV in the tens of TeV range.

We also studied the low energy constraints on the
model and found that couplings of these
 new states to the SM matter fields must be small, i.e., $\lesssim O(10^{-3})$. Interestingly if theses couplings are $<10^{-6}$ they can lead to displaced
vertices of multilepton signals for the production and decays of $E^{\pm\pm}$ at the LHC for reference masses of $M_E= 500$ and $M_S = 400$ GeV.

Other studies of doubly charged lepton contribution to a scalar diphoton resonance with different emphasis and context can be found in \cite{Dj}. For an early summary of other approaches to
the diphoton resonance, see, e.g., \cite{many}.

W.J.L. thanks Triumf and the theory department for its kind hospitality during her
visit. This  research is partially supported by the China Scholarship Council (CSC) and National
Science Foundation under Contracts No.11005033 and No.11405046. J.N.N. is partially supported by
the National Research Council of Canada through a contribution to Triumf.

\end{document}